\documentclass[10pt]{article}

\usepackage{graphics}
\input{epsf}
\usepackage{euscript,amssymb}

\font\msbm=msbm10
\def\RR{\hbox{\msbm R}}

\catcode`\@=11

\def\vereq#1#2{\lower3pt\vbox{\baselineskip1.5pt \lineskip1.5pt
\ialign{$\m@th#1\hfill##\hfil$\crcr#2\crcr\sim\crcr}}}

\def\Let@{\relax\iffalse{\fi\let\\=\cr\iffalse}\fi}
\def\vspace@{\def\vspace##1{\crcr\noalign{\vskip##1\relax}}}
\def\multilimits@{\bgroup\vspace@\Let@
 \baselineskip\fontdimen10 \scriptfont\tw@
 \advance\baselineskip\fontdimen12 \scriptfont\tw@
 \lineskip\thr@@\fontdimen8 \scriptfont\thr@@
 \lineskiplimit\lineskip
 \vbox\bgroup\ialign\bgroup\hfil$\m@th\scriptstyle{##}$\hfil\crcr}
\def\Sb{_\multilimits@}
\def\endSb{\crcr\egroup\egroup\egroup}
\def\Sp{^\multilimits@}

\newcommand{\be}[1]{\begin{equation}\label{#1}}
\newcommand{\ee}{\end{equation}}
\newcommand{\ba}[1]{\begin{eqnarray}\label{#1}}
\newcommand{\ea}{\end{eqnarray}}
\newcommand{\rf}[1]{(\ref{#1})}
\newcommand{\nn}{\nonumber}
\newcommand{\dn}{\mbox{\rm dn}}
\newcommand{\sn}{\mbox{\rm sn}}
\newcommand{\cn}{\mbox{\rm cn}}
\newcommand{\cd}{\mbox{\rm cd}}
\newcommand{\ds}{\mbox{\rm ds}}
\newcommand{\const}{\mbox{\rm const}}
\newcommand{\arcsinh}{\mbox{\rm arcsinh}}
\begin{document}

\author{Mariam Bouhmadi--L\'
opez\footnote{e-mail: mbouhmadi@imaff.cfmac.csic.es}, Pedro F.
Gonz\'{a}lez--D{\'\i}az\footnote{e-mail:
p.gonzalezdiaz@imaff.cfmac.csic.es} \\ and Alexander
Zhuk\footnote{e-mail: ai$\_$zhuk@imaff.cfmac.csic.es
\newline
on leave from: Department of Physics, University of Odessa, 2
Dvoryanskaya St., Odessa 65100, Ukraine} \\ \\ Instituto de
Matem\'{a}ticas y F\'{\i}sica Fundamental, \\ Consejo Superior de
Investigaciones Cient\'{\i}ficas,
\\ C/ Serrano 121, 28006 Madrid, Spain }

\title{Perfect fluid brane-world model}

\date{2001}
%\date{}
\maketitle

\abstract{By considering 5--dimensional cosmological models with a
bulk filled with a perfect fluid and a cosmological constant, we
have found regular instantonic solution which is free from any
singularity at the origin of the extra--coordinate and describe
5--dimensional asymptotically anti de Sitter wormhole, when the
bulk has a topology $\msbm R \times S^4 $ and is filled with dust
and a negative cosmological constant. Compactified brane-world
instantons which are built up from such instantonic solution
describe either a single brane or a string of branes. Their
analytical continuation to the pseudo--Riemannian metric can give
rise to either 4-dimensional inflating branes or solutions with
the same dynamical behaviour for extra--dimension and branes, in
addition to multitemporal solutions. Dust brane-world models with
other spatial topologies are also considered. }

\bigskip

\hspace*{0.950cm} PACS number(s): 04.50.+h, 98.80.Hw

%%%%%%%%%%%%%%%%%%%%%%%%%%%%%%%%%%%%%%%%%%%%%%%%%%%%%%%%%%%%

%%%%%%%%%%%%%%%%%%%%%%%%%%%%%%%%%%%%%%%%%%%%%%%%%%%%%%%%%%%%

\section{Introduction}
\setcounter{equation}{0}

\bigskip

Although higher dimensional cosmological models can be traced back
to the first years of gravitation theory, the idea has received a renewed
great attention in the last few years, due to the publication of pioneering
papers on brane and extra-dimension models which shed light for the solution
of fundamental physical problems including the extra-dimension
compactification and the hierarchy problem. Extra-dimensions can be
compactified, as in the
standard Kaluza-Klein theory \cite{KK}, or not, as was firstly suggested by
Akama \cite{Akama},
Rubakov and Shaposhnikov \cite{Ru-Sh} and others
\cite{Visser}.
While Akama \cite{Akama} considered the universe as a four vortex-like object
embedded in a 6--dimensional flat space-time, Rubakov and Shaposhnikov
proposed a model in (1+N)--dimensional Minkowski space-time ($N\geq 4$) where
particles were confined in a potential well, flat on the usual three spatial
dimensions and narrow along the extra-dimensions. Gravity was later on
included in 5--dimensional manifolds \cite{Visser} in order to trap particles
near a 4--dimensional Lorentzian submanifolds.

More recently, a model was proposed \cite{ADD} on
(4+N)--dimensions with $N\geq2$, where the extra-dimensions are
compact and their size, $R$, is deduced by imposing that the usual
Planck scale, $M_{Pl}$, is no longer a fundamental scale and
Planck scale in (4+N)--dimensions, $M_{Pl_{4+N}}$, is of the order
of the weak scale $M_{EW}$, $M_{Pl}^2=M_{Pl_{4+N}}^{2+N}R^N$. This
model can solve the hierarchy problem. In this framework the
gravitons can propagate in the extra--dimensions while the
standard model fields are confined to a 4--dimensional
submanifolds of thickness $M_{EW}^{-1}$ in the extra--dimensions.

Randall and Sundrum suggested a new approach to solve the
hierarchy problem \cite{RS1} by including just one extra compact
dimension. In their first model \cite{RS1}, inspired by string
theory \cite{Witten}, they considered 5--dimensional anti de
Sitter (AdS) bulk with two branes with opposite tension; our
universe is then placed on the brane with negative tension where
standard model particles are localized. In a second model
\cite{RS2}, these authors placed the universe in the brane with
positive tension, in a non compact background. In this framework,
it is possible to reproduce 4--dimensional general relativity even
if the extra--dimension is noncompact \cite{RS2}, due to the
existence of a massless gravitational bound state of Kaluza-Klein
(KK) modes which is the graviton of 4--dimensional world. While
for the noncompact case, KK spectrum is continuous without a gap,
for the compact case the KK excitations are quantized.

Branes in the Randall-Sundrum models are 4--dimensional flat
space-time and consequently, at least in principle, they can not
describe any inflationary universe. Nevertheless, as it was
pointing out by Garriga and Sasaki (GS) \cite{Gariga-Sasaki} it is
still possible to construct an inflating brane, whose geometry
corresponds to a 4--dimensional de Sitter space, surrounded by a
5-dimensional AdS. The Euclidean version of this solution can be
used for the description of the creation of the universe from
nothing. GS model has a single normalized gravitational bound
state which corresponds to the massless graviton and separates by
a gap from the massive KK modes.

Multi-brane-world models also exhibit 4--dimensional gravity
localization at the branes. In particular, a compact brane-world
model consisting of three flat branes, two with positive tensions
and an intermediate one with negative tension, embedded in
5-dimensional AdS space-time has been also considered
\cite{Santiago}. In this case, the universe is placed in one of the
positive tension branes. In addition, intersecting brane
configurations \cite{ADDK}, crystal brane-world \cite{Kaloper} and
brane periodic configuration \cite{Nam} have been investigated as
well. The massless gravitational bound state exists in all these
models.

However, most of the above mentioned brane-world models have the shortcoming
of either not including creation of inflating branes or having a singular
character for brane instantons. The main aim of the present paper is to
propose a set of singularity free 5--dimensional models which are able to
produce inflating brane-worlds. Such models should also induce localization
of 4--dimensional gravity at the branes.

The paper is organized as follows. In the next section we derive a
master equation \rf{2.9} which describes the behaviour of the
scale factor of D--dimensional instantonic (after an analytic
continuation to the Euclidean region) solutions for a precise bulk
filled with a perfect fluid and a cosmological constant,
$\Lambda_{D}$. For $D=5$, $\Lambda_{D}<0$ and spherical
4-dimensional sections, we obtain asymptotically AdS wormhole when
the bulk is filled with dust. Using this wormhole solution, in
section 3, we construct compact and noncompact brane-world
instantons. In section 4, we analyze the behaviour of the massless
gravitational KK mode for the models constructed in the previous
section. In section 5, we describe the brane-world birth from
nothing (the Euclidean solution) performing an analytical
continuation of the instantonic solution. In section 6, we
consider dust brane-world models for all the other possible
combinations of $\Lambda_{5}$ and the maximally symmetric branes
(flat, spherical, hyperbolic) that have not been considered in
section 2. Finally, in section 7, we summarize our results.

%%%%%%%%%%%%%%%%%%%%%%%%%%%%%%%%%%%%%%%%%%%%%%%%%%%%%%%%%%%%%%
\section{Multidimensional perfect fluid cosmology \label{setup}}
\setcounter{equation}{0}

\bigskip

Let us start our investigation of a multidimensional model with a
cosmological constant, $\Lambda_D$, and minimal scalar field,
$\varphi$, by writing down the action\footnote{Although the
constant minimal scalar field is really equivalent to a
cosmological term, it is convenient to single out an explicit
cosmological constant in the action. \label{1}}:
%%%%%
\ba{2.1} S &=& \frac{1}{2\kappa^2_D} \int d^{\,
D}X\sqrt{|g^{(D)}|}\left\{ R[g^{(D)}] - 2 \Lambda_D \right\}\\ &+&
\int d^{\, D}X\sqrt{|g^{(D)}|}\left\{ -\frac12 g^{(D)MN}
\partial_{M} \varphi
\partial_{N} \varphi - V(\varphi ) \right\} + S_{YGH} \, , \nn  \ea
%%%%
where $\kappa^2_D$ is D--dimensional gravitational constant and
$S_{YGH}$ the standard York-Gibbons-Hawking boundary term
\cite{YGH}. The $(D=1+d)$--dimensional metric $g^{(D)}$ is taken
to have the form
%%%%
\be{2.2} g^{(D)} = g^{(D)}_{MN}dX^M\otimes dX^N = - e^{2\gamma
(\tau )} d\tau \otimes d\tau + e^{2\beta (\tau )} g^{(d)}_{\mu
\nu} dx^{\mu} \otimes dx^{\nu}\ , \ee
%%%%
with $g^{(d)}$ the metric of d-dimensional Einstein space:
$R[g^{(d)}] = \lambda d \equiv R_d$, and in the case of constant
curvature space parameter $\lambda$ is normalized as $\lambda =
k(d-1)$, with $k=\pm 1,0$.

Consistent with the metric ansatz
\rf{2.2}, we suppose that the scalar field is also homogeneous:
$\varphi = \varphi (\tau )$. For such a scalar field energy
density and pressure are defined as:
%%%%
\be{2.3} \vphantom{\int} \begin{array}{rcl} T^0_0 &=& -\frac12
e^{-2\gamma} \dot \varphi^2 -V(\varphi )\, \equiv \; -\rho\, ,
\phantom{\frac{R^l_i}{d^l_i}} \\ T^{m_i}_{m_i} &=& \quad \frac12
e^{-2\gamma} \dot \varphi^2 -V(\varphi )\, \equiv \; P\, , \qquad
i = 1, \ldots ,d \, ,\phantom{\frac{R^l_i}{d^l_i}} \\
\end{array}
\ee
%%%%%
and supposed to satisfy the state equation:
%%%%%%
\be{2.4} P = (\alpha - 1)\rho\, . \ee
%%%%%
The conservation equation $T^M_{N;M} = 0$ has then a simple
integral
%%%%
\be{2.5} \rho = A a^{-\alpha d} \, , \ee
%%%%
where $A$ is an arbitrary constant and $a(\tau ) := \exp [\beta
(\tau )]$ is a scale factor. Obviously, $a^d$ gives the volume of
the Universe at any hypersurface $\tau = \const$ up to a spatial
prefactor $V_d = \int d^dx \sqrt{|g^{(d)}|}$. From the
energodominant condition it is usually supposed that $\rho \ge 0$
and $-\rho \le P \le \rho$, where the latter condition (which
results in $0 \le \alpha \le 2$) means that speed of sound is less
than speed of light. For the aim of generality we will not
restrict ourselves to satisfy these conditions because at least
the first of them can be violated on the case of a scalar field .
For example, $\alpha = 0$ with $\rho < 0$ describes the case of a
negative cosmological constant.

It can be shown \cite{Zhuk_QCG} that Einstein equations for such
scalar fields are equivalent to the Einstein equations for a
perfect fluid with action:
%%%%
\be{2.6} S = \frac{V_d}{\kappa^2_D} \int d \tau \left[ \frac12
e^{-\gamma + \gamma_0} d(1-d) \dot \beta^2 - e^{\gamma - \gamma_0}
U \right] \equiv \frac{V_d}{\kappa^2_D} \int d \tau L \, , \ee
%%%%
in which we performed integration over spatial coordinates $x$
($V_d$ is an artifact of this integration), $\gamma_0 := d\beta$,
overdot denotes differentiation with respect to $\tau$ and we have
introduced a "potential energy"
%%%%%
\be{2.7} U = e^{2 \gamma_0} \left( -\frac12 R_d e^{-2\beta} +
\Lambda_D + \kappa^2_D\, \rho \right)\, .\ee
%%%%%
The constraint equation corresponding to the Lagrangian in
Eq.\rf{2.6} results in the following master equation:
%%%%%
\be{2.8} \frac{\partial L}{\partial \gamma} = 0 \quad
\Longrightarrow \quad \frac12 e^{-\gamma + \gamma_0} d(1-d) \dot
\beta^2 + e^{\gamma - \gamma_0} U = 0 \, , \ee
%%%%
which is exactly the 00--component of the Einstein equation.
Function $\gamma$ reflects the freedom for the choice of different
time gauges: $\gamma = 0$ is the proper time gauge, $\gamma =
\gamma_0$ is the harmonic time gauge \cite{IMZ} and $\gamma =
\beta$ is the conformal time gauge. Thus, in the proper time gauge
($\gamma = 0$) Eq.\rf{2.8} reads
%%%%%%
\be{2.9} \left(\frac{\dot a}{a}\right)^2 + \frac{R_d}{d(d-1)}
\frac{1}{a^2} + \frac{2}{d(1-d)} \Lambda_D + \frac{2}{d(1-d)}
\kappa^2_D A\, a^{-\alpha d} = 0 \, .\ee
%%%%%

We will concentrate mainly on 5--dimensional brane-world models with
AdS bulk\footnote{5--D AdS solutions of the Einstein equations are
usually motivated by M--theory \cite{Witten}. However,
studying multidimensional cosmological brane-world solutions with
positive bulk cosmological constant \cite{hep-th/0108109} is also
of interest and we shall consider models with non-negative
bulk cosmological constant in section 6. \label{2}}: $D = 5, \,
\Lambda_D = \Lambda_5 \equiv -|\Lambda_5| <0$.
The scalar curvature $R_d$ can be an arbitrary
constant but we shall consider the particular case of constant
curvature 4-D space $R_d = k d(d-1) = 12k$, restricting mainly to
the positive curvature with $k = +1$. For these parameters
Eq.\rf{2.9} becomes
%%%%%
\be{2.10} (\dot a)^2 + k + \Lambda a^2 - \bar A^2 \, a^{-4\alpha +
2} = 0\, , \ee
%%%%
where for simplicity we have introduced the following notation:
$|\Lambda_5| /6 \equiv \Lambda$ and $(1/6) \kappa^2_5 A \equiv
\bar A^2$. Some of the particular values $\alpha$ are of special
interest because they correspond to important types of matter. For
example, in 5--D space-time\footnote{See e.g.
\cite{asrto-ph/9908047} and references therein for the discussion
of the equations of state for different kinds of cosmic defects in
usual 4--D universe. In these references a regime where the number
of defects per co-moving volume is a constant is considered.
\label{3}} $\alpha = 5/4$ describes radiation, $\alpha = 1$
corresponds to dust (0--D objects), $\alpha = 3/4$ represents
cosmic strings (1--D objects), $\alpha = 1/2$ describes domain
walls (2--D objects), $\alpha = 1/4$ corresponds to hyperdomain
walls (3--D objects) and $\alpha = 0$ represents vacuum (which can
be in some sense considered as a 4--D object).

Now, as we want to construct 5--D brane-world models, we perform
the Wick rotation to Euclidean "time" $r:\, \tau \to - ir$. Where
$r$ is to be considered as an extra spatial coordinate orthogonal
to 4--D branes, i.e., now hypersurfaces at $r = const$. Then, in
proper "time" gauge, metric \rf{2.2} is
%%%%
\be{2.11} g^{(5)} = dr \otimes dr + a^2(r) g^{(4)} \, \ee
%%%%
and the Euclidean version of Eq.\rf{2.10} reads
%%%%
\be{2.12} (\dot a)^2 -k -\Lambda a^2 + \bar A^2 \, a^{-4\alpha +
2} = 0\, .\ee
%%%%
Solutions of this equation describe 5--D instantons which can be
used to construct brane-world models. Obviously, the precise form
of the resulting instantons depends on the type of perfect fluid
we use, i.e., on the choice of the parameter $\alpha$. The vacuum
case $\alpha = 0$ (which here corresponds to simple redefinition
of the cosmological constant: $\Lambda_D + \kappa^2_D A \to
\Lambda_D$) was considered in paper \cite{Gariga-Sasaki} for
positive curvature $k =+1$ with solution
%%%%
\be{2.13} a(r) = l\, \sinh (r / l)\, , \ee
%%%%
where $l := \sqrt{\Lambda^{-1}}$ is the AdS radius.

In the present paper we investigate the case of dust ($\alpha =
1$) with $k =+1$. For this particular case the solution of
Eq.\rf{2.12} reads:
%%%%%
\be{2.14} a(r) = \frac{l}{\sqrt{2}} \left( \sqrt{b}\, \cosh
\frac{2r}{l} -1 \right)^{1/2}\, , \qquad -\infty < r < + \infty \,
, \ee
%%%%
where $b := 1 + 4 \Lambda \bar A^{2}$. It can be easily seen that
this solution is reduced to \rf{2.13} in the limit: $\bar A^{2}
\to 0 \Longrightarrow b \to 1$. Metric \rf{2.11} with solution
\rf{2.14} describes a wormhole (the integration constant being
taken in such a way that $r=0$ corresponds to the wormhole
throat).

It is useful to present solution \rf{2.14} in the conformal "time"
gauge: $dr = \pm a d \eta$. It can be easily seen that
%%%%
\be{2.15} a(\eta ) = l\, b^{1/4}\; \frac{\dn\, b^{1/4} |\eta
|}{\sn\, b^{1/4} |\eta |}\; , \qquad 4 n K (m)\, \le\, b^{1/4}
|\eta | \, \le \, 2(1+2n) K (m)\; ,\quad n = 0, 1, 2, \ldots \quad
,\ee
%%%%
satisfies the corresponding equation
%%%%
\be{2.16} \left( \frac{d a}{d \eta} \right)^2 - a^2 -\Lambda a^4 +
\bar A^2 = 0 \, . \ee
%%%%
Functions $\dn\, u \equiv \dn \,(u|m)$ and $\sn\, u \equiv \sn \,
(u|m)$ (with $m = (b^{1/2} +1)/2b^{1/2}$ ) are Jacobian elliptic
functions and $K (m) = F (\pi/2 | m)$ is the complete elliptic
integral of the first kind \cite{Abramowitz}. For each of the
coordinate intervals, defined by the choice of $n$ and sign of
$\eta$, there is an one--one correspondence between solution
\rf{2.15} and wormhole \rf{2.14} : $b^{1/4}\; |\eta | \to 4 n K(m)
\Longrightarrow r \to + \infty \; , \, b^{1/4}\; |\eta | \to
2(1+2n) K(m) \Longrightarrow r \to -\infty \, $, and $\ b^{1/4}\;
|\eta | = (1+4n)K(m) \Longrightarrow r = 0$ is the wormhole
throat. Coordinates $r$ and $\eta$ are connected with each other
by :
%%%%
\be{2.17} b^{1/4}\; |\eta | = F \left(\, \varphi \, |\, m\,
\right) = u \quad \Longleftrightarrow \quad \sn\; b^{1/4} |\eta |
= \frac{1}{\cosh r/l} \; ,\ee
%%%
where $\varphi := \arcsin (1/\cosh (r/l))$ and $F \left(\, \varphi
\, |\, m\, \right)$ is the incomplete elliptic integral of the
first kind. In the limit $\bar A^{2} \to 0 \Longrightarrow b \to 1
\Longrightarrow m\to 1\, $, functions $\dn \, (u|m) \to 1/\cosh
|\eta | \, , \; \sn \, (u|m) \to \tanh |\eta | \, $ and solution
\rf{2.15} tends to $a(\eta ) \to l/ \sinh |\eta |$, which in fact
is solution \rf{2.13} as expressed in the conformal "time" gauge.

%%%%%%%%%%%%%%%%%%%%%%%%%%%%%%%%%%%%%%%%%%%%%%%%%%%%%%%%%%%%%%
\section{Brane-world instantons\label{instanton}}
\setcounter{equation}{0}

\bigskip

%%%%%%%%%%%%%%%%%%%%
In this section, we use solutions \rf{2.14} and \rf{2.15} to
construct brane-world instantons. This can be performed by
excising regions with $r > L$ for two identical wormholes
\rf{2.14} and gluing the remaining two copies along the two
4-spheres at $r=L$. The obtained instanton can be described by the
following piecewise continuous function:
%%%%
\be{3.1} a (r) = \left\{\begin{array}{rcl} \frac{l}{\sqrt{2}}
\left[ \sqrt{b}\, \cosh \left(\frac{2r}{l}\right) -1
\right]^{1/2}\qquad &\, , &\quad -\infty < r \le L \\
\frac{l}{\sqrt{2}} \left[ \sqrt{b}\, \cosh
\left(\frac{2(2L-r)}{l}\right) -1 \right]^{1/2} &\, , & \qquad L
\le r < +\infty
\\
\end{array}\right. \ee
%%%%
This function is continuous but not smooth at the gluing "point"
$r=L$. Due to the Lanczos-Israel junction condition, this results
in the appearance of a 4--D spherical brane at $r=L$ with a
tension:
%%%%%%
\be{3.2} T(r=L)\, =\, \frac{1}{\kappa_5^2}\frac{3}{4} \widehat K
(L)\, =\, \frac{6}{\kappa_5^2 l}\; \frac{\sqrt{b}\, \sinh
(2L/l)}{\sqrt{b}\, \cosh (2L/l) - 1} =\, \frac{6}{\kappa_5^2 l}\;
\frac{\sinh (L/l)\, \cosh (L/l)}{\sinh^2 (L/l) + m_1}\, > \, 0\, ,
\ee
%%%%%
where $m_1 = 1-m\, ,\; \widehat K (L) \equiv K (L^+) - K (L^-)$,
and $K(L) = -4 a^{-1}(da/dr)_{r=L}$ is the trace of the extrinsic
curvature (for the case of 5--D metric \rf{2.11} written in the
form of the Gaussian normal coordinates).

Instantonic solution \rf{3.1} is non-compact: the scale factor
$a(r)$ goes to $\infty$ when $r \to \pm \infty$, and cannot be
used for the description of the brane-world birth from "nothing"
(from the Euclidean region) because the probability of it in this
case is equal to zero. On the other hand, as we shall see below,
the bound state of the spin--2 gravitational perturbations
(corresponding to the Newtonian gravity on the brane) is
proportional to $a^{3/2}$ and is divergent when $a \to \infty$.
However, we can compactify this instanton identifying points
corresponding to the throats at $r=0 \leftrightarrow r=2L$. It
changes the topology along the extra dimension from $\mathbb{R}$
to $S^1$. Now, the range of the variation of $r$ is the interval
$[0,\, 2L]$. Since the geometry is smoothly glued at these points,
such procedure does not lead to the appearance of new
branes\footnote{Instanton \rf{3.1} can also be compactified by
additional cutting at distances $\triangle r=L_1 < L$ from the two
sides of the brane, gluing then along these two cuts. Geometry is
not smoothly matched at this surface which results in a new
negative tension brane. However, in our paper we will not consider
such a compact instanton because the first one, with the wormhole
throat identification, seems to us more elegant and it leaves a
possibility for 5--D baby-universe nucleation at the throat.
\label{4}}.

This procedure for the construction of the brane-world instanton
can be easily generalized to the case of an arbitrary number of
parallel branes by gluing one-brane manifolds at throats and
identifying the two final opposite throats. For example, in the
case of $n$ branes, located at the distances $r = L_i\, ,\; i= 1,
\ldots , n$ from throats, this instanton can be described by the
following piecewise continuous function:
%%%%%
\ba{3.3} a(r)& = & \sum_{i=1}^{n+1} a_i (r) \theta_i (r)\; ,
\qquad \qquad \qquad 0\, \le \, r \, \le \, 2 \sum_{i=1}^n L_i \;
,\\
 a_i(r)& = & \frac{l}{\sqrt{2}} \left[ \sqrt{b}\, \cosh
\left(\frac{2(2\sum_{k=1}^{i-1} L_k -r)}{l}\right) -1
\right]^{1/2}\, , \quad L_{i-1} +2\sum\limits_{k=1}^{i-2}L_k\; \le
\, r\, \le \; L_{i} +2\sum\limits_{k=1}^{i-1} L_k\; ,\nn \ea
%%%%%
where $L_0 \equiv L_{n+1} \equiv 0$ and
%%%%%%
\be{3.4} \theta_i (r) =\tilde \eta (r-r_{i-1} ) -\tilde \eta
(r-r_i) =\left\{\begin{array}{rcl} &0&\, , \quad r < r_{i-1}\\
&1&\, , \quad r_{i-1} \le r < r_i \\ &0&\, , \quad r \ge r_i \\
\end{array}\right. \ee
%%%%%
are piecewise discontinuous functions, with $\tilde \eta (r -r_i)$
being the usual step function\footnote{ We put a tilde above it to
distinguish from the conformal time $\eta$. Obviously, $
\theta_i^{\, p} = \theta_i\; ,\; p>0\; ;\quad \theta_i \, \theta_j
= 0 \; ,\; i \ne j\; \Longrightarrow a^p = \sum_{i=1}^n a_i^p
\theta_i \; ,\; \forall p$ and $ \theta_i^{'} = \delta (r-r_{i-1})
- \delta (r-r_i)$. \label{5}} equal to zero for $r<r_i$ and
becoming unity for $r \ge r_i$. We have redefined the coordinate $r$ in
such a way to cover the range of variable for our solution by one
coordinate chart. In this case the i-th brane has coordinate $r=
r_i = L_i + 2\sum_{k=1}^{i-1}L_k\, , \; i = 1, \ldots ,n$ and the
i-th throat is located at $r= r_{(th)i} = 2\sum_{k=1}^{i-1}L_k\, ,
\; i = 1, \ldots ,n+1$. "Points" $r =r_0 \equiv r_{(th)1} =0$ and
$r = r_{n+1} \equiv r_{(th)n+1} = 2 \sum_{i=1}^n L_i$ are
identified with each other due to $S^1$-symmetry. Thus, metric
\rf{2.11} with scale factor \rf{3.3} describes a 5--D compact
instanton with $n$ parallel branes transversal to coordinate $r$
(see figure 1). Each of these branes has a tension given by
Eq.\rf{3.2} with the evident substitution $L \to L_i$ for the i-th
(i=1,\ldots ,n) brane.

The similar procedure for construction of brane-world instanton
can be performed in the conformal gauge. For example, one-brane
instanton can be obtained from the solution \rf{2.15} if we take
two wormholes with $\eta \lessgtr 0$ and $n=0$, cut them at $|\eta
| = \eta_0 < b^{-1/4} K(m)$, excise regions $|\eta | < \eta_0$ and
glue them along this cut. The obtained brane-world instanton is:
%%%%
\be{3.5} a(\eta ) = l\, b^{1/4}\; \frac{\dn\, b^{1/4} (|\eta | +
\eta_0) }{\sn\, b^{1/4} (|\eta | +\eta_0)}\; ,\ee
%%%%
where we redefined the coordinate $\eta $ in such a way ( $|\eta
|_{old} \to |\eta |_{new} + \eta_0$ ) to cover this instanton by
one coordinate chart. Here, $0 \le b^{1/4} |\eta | \le 2K(m) -
b^{1/4} \eta_0$ and $0 \le b^{1/4} |\eta | \le K(m) - b^{1/4}
\eta_0$ (with the identification $-b^{-1/4} K(m) +\eta_0
\leftrightarrow b^{-1/4} K(m) -\eta_0$) correspond respectively to
non-compact and compact instantons with brane at $\eta =0$ and
tension
%%%%
\be{3.6} T(\eta =0)\, =\, \frac{6}{\kappa_5^2 l}\; \frac{\cn\,
\bar \eta_0}{\dn^2\, \bar \eta_0} =\, \frac{6}{\kappa_5^2 l}\;
\frac{\sqrt{1 - \sn^2\, \bar \eta_0}}{1 - m\; \sn^2\, \bar
\eta_0}\,
> \, 0\, , \ee
%%%%%
which coincides with Eq.\rf{3.2} if $\eta = \eta_0$ and $r=L$ are
connected by formula \rf{2.17}. In Eq.\rf{3.6} we use the
shortage: $\bar \eta \equiv b^{1/4} \eta$ which we shall employ
below. It is necessary to note that the distance between the
throat and the brane is: $\bar \xi_0 = K(m) - \bar \eta_0$ and
parameter $\xi_0$ becomes indefinite in the limit $m \to 1$
because in this case\footnote{Here, the wormhole pinches off
because there is no matter in this limit. \label{6}} $K(m) \to
\infty$. In term of parameter $\xi_0$, tension \rf{3.6} reads:
%%%%
\be{3.7} T \, =\, \frac{6}{\kappa_5^2 l}\; \frac{1}{\sqrt{m_1}}\;
\, \sn\, \bar \xi_0\; \, \dn\, \bar \xi_0 =\, \frac{6}{\kappa_5^2
l}\; \frac{1}{m_1}\, \dn\, \bar \xi_0\, \sqrt{\dn^2\, \bar \xi_0 -
\cn^2\, \bar \xi_0}\, \, , \ee
%%%%%
which, of course, coincides with expression \rf{3.2}, as it can be
checked using the relations $\sn^2 \, \bar \xi_0 = \sinh^2
\frac{L}{l} /(\cosh^2 \frac{L}{l} - m)\, , \; \cn^2 \, \bar \xi_0
= m_1 /(\cosh^2 \frac{L}{l} - m)\,$ and $\dn^2 \, \bar \xi_0 = m_1
\; \cosh^2 \frac{L}{l} /(\cosh^2 \frac{L}{l} - m)$.

In this gauge, the compact $n$-brane instanton (with branes at
distances $\eta = \xi_i ,\; i=1,\ldots ,n$ from the throats) can
be written in a form similar to expression \rf{3.3}:
%%%%%
\ba{3.8} a(\eta)& = & \sum_{i=1}^{n+1} a_i (\eta ) \theta_i (\eta
)\; , \qquad \qquad \qquad 0\, \le \, \eta \, \le \, 2
\sum_{i=1}^n \xi_i \; ,\\
 a_i(r)& = & l\, b^{1/4}\; \ds\, \left( \bar \eta +
(4i-3)K(m) - 2 \sum_{k=1}^{i-1}\bar \xi_k \; |\; m \right)  \, ,
\quad \xi_{i-1} +2\sum\limits_{k=1}^{i-2}\xi_k\;
\le \, \eta \, \le \; \xi_{i} +2\sum\limits_{k=1}^{i-1} \xi_k\;
,\nn \ea
%%%%%
where $\ds(u|m)\equiv\dn(u|m)/\sn(u|m)$,  $\xi_0 \equiv \xi_{n+1}
\equiv 0$ and
%%%%%%
\be{3.9} \theta_i (\eta ) =\tilde \eta (\eta -\eta_{i-1} ) -\tilde
\eta (\eta -\eta_i) =\left\{\begin{array}{rcl} &0&\, , \quad \eta
< \eta_{i-1}\\ &1&\, , \quad \eta_{i-1} \le \eta < \eta_i \\ &0&\,
, \quad \eta \ge \eta_i
\\
\end{array}\right. \ee
%%%%%
It can be easily seen that \rf{3.8} is a piecewise continuous
function. We have also redefined the coordinate $\eta $ to cover
our solution by one coordinate chart. In this case the i-th brane
has coordinate $\eta = \eta_i = \xi_i + 2\sum_{k=1}^{i-1}\xi_k\, ,
\; i = 1, \ldots ,n$\, , the throats have coordinates $\eta =
\eta_{(th)i} = 2\sum_{k=1}^{i-1}\xi_k\, , \; i = 1, \ldots ,n+1$
and "points" $\eta =\eta_0 =0$ and $\eta = \eta_{n+1} = 2
\sum_{i=1}^n \xi_i$ are identified with each other due to the
$S^1$-symmetry. Each of the branes has tension with the form given
by Eq.\rf{3.7}, with an evident substitution $\xi_0 \to \xi_i$ for
the $i$-th (i=1,\ldots ,n) brane.

%%%%%%%%%%%%%%%%%%%%%%%%%%%%%%%%%%%%%%%%%%%%%%%%%%%%%%%%%%%%%%
\section{Volcano and Kurile-ridge potentials\label{potential}}
\setcounter{equation}{0}

\bigskip

%%%%%%%%%%%%%%%%%%%%
It is well known for metrics of the form \rf{2.2} and \rf{2.11},
rewritten in the conformal time gauge, that if tensor metric
perturbations $\delta g_{MN} (X) = h_{MN} (\eta ,x)$ is taken in
the gauge: $h_{\eta \eta} = h_{\eta \mu} = h_{\mu}^{\mu} ={h^{\mu
\nu}}_{;\nu} = 0$ (where the semicolon denotes a covariant
derivative with respect to d--dimensional metric $g_{\mu
\nu}^{(d)}(x)$) and is normalized and factorized as $h_{\mu \nu}
(\eta , x) = a^{-(d-5)/2}(\eta ) \psi (\eta )\hat h_{\mu \nu}(x)$,
then the radial profile of the perturbations satisfies a
Schr$\ddot{o}$dinger-like equation \cite{RS2}
%%%%
\be{4.1} \left( -\frac{d^2}{d\eta^2} + V(\eta ) \right) \psi (\eta
) = M^2 \psi (\eta )\, , \ee
%%%%
in which $V(\eta)$ is the so-called "volcano" potential
%%%%
\be{4.2} V(\eta ) =\left. \left( \frac{d^2}{d\eta^2}
a^{\frac{d-1}{2}}\right) \right/ a^{\frac{d-1}{2}}\, , \ee
%%%%
and $M$ defines a mass spectrum of spin--2 perturbations $\hat
h_{\mu \nu}(x)$ moving in the d--dimensional background metric
$g^{(d)}_{\mu \nu}(x)$. If the spectrum of mass starts from $M=0$,
then a state corresponding to this value of $M$ describes usual
4--dimensional (for $d = 4$) massless gravitons responsible for
the standard Newtonian gravity.

It can be seen that for $M=0$ the trivial solution of Eq.\rf{4.1}
is
%%%%
\be{4.3} \psi \propto a^{\frac{d-1}{2}} \, .\ee
%%%%
Thus, if the scale factor $a$ has maxima at branes, this zero-mode
state is localized at the branes. On the other hand, if the shape
of the volcano potential is such that it suppresses the
Kaluza--Klein modes with $M>0$ at the branes by the wings of the
potential, then the massless graviton gives the main contribution
to gravity on the brane and effectively we have the Newtonian
gravitation here. Moreover, to have a well defined effective
gravitation on branes, it is desirable to have a mass spectrum
where the massless graviton is well separated by a mass gap from
other massive Kaluza--Klein modes \cite{Nam}. This is precisely
the general situation that we have for all models we consider in
the present paper, as we are going to show in what follows.

Let us first consider the one-brane case described by Eqs.\rf{3.5}
- \rf{3.7} (or \rf{3.1} and \rf{3.2}). The scale factor $a$,
defined by Eq.\rf{3.5}, has a local maximum at the brane $\eta =
0$, then tends to its minima at the throat positions $|\bar \eta |
= K(m) -\bar \eta_0$, and diverges at the limit $|\bar \eta | \to
2K(m) -\bar \eta_0$. Thus, for the non-compact model the radial
profile $\psi \propto a^{3/2}$ is delocalized in this limit.
However, in the compact case ( with identification of the throats)
the local maximum at the brane becomes the global one (see figure
2-a) and $\psi$ provides the desired zero-mode bound-state. The
volcano potential reads in this case:
%%%%
\ba{4.4} V(\eta )&=& \frac{15}{4} b^{1/2} \frac{1}{(\sn \, (|\bar
\eta | + \bar \eta_0 ))^2} - \frac{3}{4} b^{1/2} \left[ 3 -
m\left( \frac{\cn \,(|\bar \eta | + \bar \eta_0 )}{\dn \, (|\bar
\eta | + \bar \eta_0 )}\right)^2 \right] - 3 b^{1/4} \frac{\cn \,
(|\bar \eta | + \bar \eta_0 )}{\dn \, (|\bar \eta | + \bar \eta_0
)\; \sn \, (|\bar \eta | + \bar \eta_0 )} \delta (\eta ) \nn \\
&\left. \right.& \\ &=&\frac{15}{4} b^{1/2} \left( \frac{\dn \,
(\bar \xi_0 - |\bar \eta |) }{\cn \, (\bar \xi_0 - |\bar \eta |)}
\right)^2 - \frac{3}{4} b^{1/2} \left[ 3 - m \left( \sn \, (\bar
\xi_0 - |\bar \eta |)\right)^2 \right] - 3 b^{1/4} \frac{\sn \,
(\bar \xi_0 - |\bar \eta |)\; \dn \, (\bar \xi_0 - |\bar \eta
|)}{\cn \, (\bar \xi_0 - |\bar \eta |)} \delta (\eta ).\nn\ea
%%%%
Thus, at the brane, $\delta$--function provides the zero-mode
localization and the wings of the potential clearly provide the
suppression of the Kaluza--Klein modes: the wings approach maximum
at the brane $\overline V(\eta = 0) \equiv V_{br}$ (where
$\overline V(\eta )$ is the truncated potential \rf{4.4} without
the $\delta$--function term) and reach their minimum at the throat
(see figure 2-b) :
%%%%
\be{4.5} \left. V(\eta )\right|_{|\bar \eta | = K(m) -\bar \eta_0
= \xi_0}\; \equiv \, V_{th}\, =\, \left\{\begin{array}{rcl} (3/2)
\sqrt{b} &\, , &\quad b \ne 1\\ 9/4\quad &\, , & \quad b = 1
\\
\end{array}\right. \ee
%%%%
Two different values of the minimum result from the two different
asymptotic behaviours of the function $\cd\,(u | m)\equiv\cn \,(u
| m) / \dn \, (u | m) \to 0 /\sqrt{1-m}$ when $ u \to K(m)$.
Hence, this fraction equals to 0 if $m \ne 1$ and goes to $1$ when
$m \to 1$. The second value of Eq.\rf{4.5} $V_1 = 9/4$ was
obtained in \cite{Gariga-Sasaki} for model \rf{2.13}. Thus,
Kaluza--Klein modes starts at
%%%%
\be{4.6} M = \sqrt{V_{th}}\, , \ee
%%%%
and this mass gap separates the zero-mode from other massive
Kaluza--Klein modes\footnote{It is clear, that to an observer on a
i-th brane at $r=r_i$, the physical metric is the induced metric
on this brane: $g^{(4)}_{(ph)\mu \nu} = a_i^2(r_i ) g^{(4)}_{\mu
\nu}$. It results in an appropriate rescaling of effective
physical values of, for example, effective 4--D cosmological and
gravitation constants on the brane (see \cite{Bouhmadi-Zhuk}).
Obviously, physical Kaluza--Klein masses for this observer are
similarly rescaled, $M \to m_{(ph)} = M/a_i(r_i )$. \label{7}}. It
is obvious that the spectrum of the Kaluza--Klein modes is
discrete for the compact models.

In the case of the compact n-brane models described by \rf{3.8}
the radial profile reads:
%%%%
\be{4.7} \psi \propto a^{3/2}(\eta )= \sum_{i=1}^{n+1} a_i^{3/2}
(\eta ) \theta_i (\eta )\, ,\ee
%%%%
where we used the properties of the $\theta$--function (see
footnote \ref{5}). This formula shows that the massless graviton
(zero-mode) is localized on each of the branes as they approaches
their local maximum at the branes (see Eq.\rf{3.8}).

The volcano potential \rf{4.2} in this case can be written in a
more compact form in the proper "time" gauge
%%%%
\ba{4.8} V &=& a^{-3/2} \frac{d^2}{d\eta^2} a^{3/2} = a^{-3/2}
\left( a \frac{d a}{d r} \frac{d }{d r} a^{3/2} + a^2 \frac{d^2}{d
r^2} a^{3/2}\right)\nn \\ &=& \sum_{i=1}^{n+1} \overline
V_i(r)\theta_i (r)- \frac{16}{3} \kappa^2_5 \sum_{i=1}^{n} a_i^2 T
(r) \delta (r - r_i)\, , \ea
%%%%
where
%%%%
\ba{4.9} \overline V_i(r) &=& \frac{9}{4} \left(\frac{d a_i}{d
r}\right)^2 + \frac{3}{2} a_i \frac{d^2 a_i}{d r^2}\\ &=&
\frac{15}{4}\sqrt{b} \cosh^2 (r-2\sum_{k=1}^{i-1}L_k) / l\; \;
\frac{\sinh^2 (r-2\sum_{k=1}^{i-1}L_k)/l +(4/5)m_1}{\sinh^2
(r-2\sum_{k=1}^{i-1}L_k)/l + m_1} \, -\, \frac{3}{2}\sqrt{b}\nn \;
. \ea
%%%%
%\\ &=& \frac{9}{4}\sum_{i=1}^{n+1}
%\left(\frac{d a_i}{d r}\right)^2\theta_i +
%\frac{3}{2}\sum_{i=1}^{n+1} a_i \frac{d^2 a_i}{d r^2} \theta_i -
%\frac{16}{3} \kappa^2_5 \sum_{i=1}^{n} a_i^2 T (r) \delta (r -
%r_i)\, , \nn \ea
%%%%
Here, we used the expressions \rf{3.3} and \rf{3.4} for the
functions $a_i$ and $\theta_i$ and the tension $T(r_i)$ has the
form of Eq.\rf{3.2} with the replacement $L \to L_i\, , \; i =
1,\ldots ,n$.
%where functions $a_i$ and $\theta_i$ are defined in
%\rf{3.3} - \rf{3.4} and tension $T(r_i)$ has the form of \rf{3.2}
%with the substitution $L \to L_i\, , \; i = 1,\ldots ,n$.
It can be seen that potential \rf{4.8} has its minima at the
throats, $\left. V(r) \right|_{r_{(th)i}} = V_{th}\, , \; i=
1,\ldots ,n\, $, defined by expression \rf{4.5}, and reaches its
local maxima at the branes,
%%%%
\be{4.10} V_{(br)i} = \frac{15}{4}\sqrt{b} \cosh^2 (L_i/l)
\frac{\sinh^2 (L_i/l) +(4/5)m_1}{\sinh^2 (L_i/l) + m_1} \, - \,
\frac{3}{2}\sqrt{b}\; , \quad i=1,\ldots ,n \, , \ee
%%%%
which tend to $ (15/4) \cosh^2 (L_i/l) - 3/2$ in the limit $b\to
1$, corresponding to the vacuum case \rf{2.13}.

Thus, in the n-brane case, potential \rf{4.8} has the form of a
string (closed string for compact models) of volcano potentials
(see figure 3) and can be named as "Kurile-ridge potential".
Obviously, this form of the potential entails suppression of the
massive Kaluza-Klein modes at the branes.

To conclude this section, we would like to remark an interesting
possibility arising from the special form of the Kurile-ridge
potential with periodic structure. In this case we obtain a band
structure for the mass spectrum of the Kaluza-Klein modes and only
masses from these bands are allowed to exist (see e.g.
\cite{Nam}). This would give rise to separation between the
zero-mode and the massive Kaluza--Klein modes even larger
than those induced by the mass gap \rf{4.6}.

%%%%%%%%%%%%%%%%%%%%%%%%%%%%%%%%%%%%%%%%%%%%%%%%%%%%%%%%%%%%%%
\section{Brane-world birth from "nothing"\label{birth}}
\setcounter{equation}{0}

\bigskip

%%%%%%%%%%%%%%%%%%%%
In this section we investigate the possibility for the creation of
 brane worlds from "nothing". To be more precise, we interpret
the brane-world instantons described in section 3 as a
semiclassical paths for the quantum tunneling from the Euclidean
region ("nothing").

Euclidean metric \rf{2.11} was obtained from the Lorentzian one by
the Wick rotation $\tau \to -ir$. For the compact solutions of
section 3 this Euclidean metric describes the compact 5--D
brane-world instantons. It is clear that back Wick rotation $r \to
i\tau$ will destroy these branes leaving instead of the brane a
solution corresponding to 5--D baby universe branching off from
the wormhole throats at $r=0$. This baby universe represents 5--D
FRW-like universe filled with dust and a negative cosmological
constant. It has a maximum at time corresponding to the wormhole
throat and bounces from a minimum occurring when the scale factor
equals to zero (usual FRW-type singularity). However, there is a
possibility for the analytical continuation to the Lorentzian
space-time which preserves branes. As it was printing out in paper
\cite{Gariga-Sasaki}, it can be done for the case when 4--D
positive curvature metric $g^{(4)}$ in Eq.\rf{2.11} is the metric
of a 4--sphere:
%%%%
\be{5.1} ds^2_E = dr^2 + a^2(r)(d\chi^2 + \sin^2\chi\,
d\Omega^2_{(3)})\, .\ee
%%%%
Then, the analytic continuation of the azimuthal coordinate
$\chi$:
%%%%
\be{5.2} \chi \longrightarrow iHt + \frac{\pi}{2}\, , \ee
%%%%
results in a Lorentzian metric with evolving branes:
%%%%
\be{5.3} ds^2_L = dr^2 + H^2 a^2(r)( - dt^2 + \frac{1}{H^2}\cosh^2
Ht \; d\Omega^2_{(3)})\, . \ee
%%%%
Here, parameter $H$ is chosen in such a way that $t$ describes the
proper time on the brane. For example, in the one-brane-case
\rf{3.1}:
%%%%
\be{5.4} \left. H a(r)\right|_{r=L} = 1 \quad \Longrightarrow
\quad H = \frac{1}{a(L)} = \frac{\sqrt{2}}{l} \left( \sqrt{b}\,
\cosh \frac{2L}{l} -1 \right)^{-1/2}\, .\ee
%%%%
The compact version of solution \rf{3.1} describes a 4--D
spherical brane of radius $a(L) = H^{-1} $ enclosing 5--D space
with frozen dust and negative cosmological constant (AdS bulk).
The south pole of the 4--D sphere with the azimuth coordinate
$\chi = 0$ can be treated as "nothing". The birth of the
brane-world takes place at the time $t=0$ which corresponds in the
Euclidean region to $\chi = \pi /2$ (the equator of the 4--D
sphere) where the sections $\chi = \const$ of the 4--D sphere
reach their maximum with radius $H^{-1}$.
%when the "radial" coordinate on 4--D sphere ($\rho =
%\sin \chi$) reaches its maximum.
After birth, the brane represents an evolving 3--D sphere with
initial radius $H^{-1}$. More precisely, it is an inflating de
Sitter space with the Hubble constant $H^{-1}$ (see figure 4).

In the case of $n$ branes described by expression \rf{3.3}, we can
introduce for each of the coordinate patches between the throats
$r_{(th)i} \le r \le r_{(th)i+1}$ the following transformation:
%%%%
\be{5.5} \chi \longrightarrow iH_it_i + \frac{\pi}{2}\, , \ee
%%%%
where
%%%%
\be{5.6} H_i = \frac{1}{a(r_i)} \ee
%%%%
and $r_i \in [r_{(th)i}, r_{(th)i+1}]$ is the position of the i-th
brane ($i=1,\ldots ,n$). After this continuation, the Lorentzian
metric is:
%%%%
\be{5.7} ds^2_L = dr^2 + H_i^2 a^2(r)( - dt_i^2 +
\frac{1}{H_i^2}\cosh^2 H_it_i \; d\Omega^2_{(3)})\, ,\ee
%%%%
where coordinates $t_i$ are glued to each other at the throats:
%%%%
\be{5.8} t_i = \frac{a_i}{a_j}t_j\, . \ee
%%%%
Thus, each of $n$ branes at $r=r_i$ represents a de Sitter space
with its own proper time $t_i$ and own Hubble constant $H_i$.

To conclude this section, we would like shortly comments some alternative
possibilities of the analytic continuation. As we have noted
before, the analytical continuation $r\to i\tau$ leads to the
creation of a 5--D baby universe with scale factor,
%%%%
\be{5.9} a(\tau)= \frac{l}{\sqrt{2}}\left(\sqrt{b}
\cos\frac{2\tau}{l} - 1\right)^{1/2}\; , \quad 0 \le \tau \le
\frac{l}{2} \arccos \frac{1}{\sqrt{b}}\, . \ee
%%%%
>From this model we can actually construct a spherical 3--D brane
by using a similar procedure to that we have described in section
3; that is excising regions with $\chi > L < \pi $ for two
identical 4-sphere and gluing the remaining two copies along
3--spheres $\chi = L$. The brane-world model in this case is
described by the metric:
%%%%
\be{5.10} ds^2_L = -d\tau^2 + a^2(\tau)( d\chi^2 + a^2(\chi )\;
d\Omega^2_{(3)})\, ,\ee
%%%%
where the scale factor $a(\tau )$ is defined by Eq.\rf{5.9} and
the scale factor $a(\chi )$ reads
%%%%
\be{5.11} a (\chi ) = \left\{\begin{array}{rcl} \sin \chi \quad
\quad \quad &\, , &\quad 0 \le \chi \le L < \pi
\\ \sin (2L-\chi) &\, , & \quad L \le \chi \le 2L
\\
\end{array}\right. \ee
%%%%
In this model, both the additional space as well as 3--D brane
have the same dynamical behaviour which is described by
Eq.\rf{5.9}.

We turn finally to another interesting possibility consisting in
the simultaneous analytic continuation in two directions: $\chi
\to it + \pi/2$ and $r \to i\tau$. It results in a metric:
%%%%
\be{5.12} ds^2_L = -d\tau^2 + a^2(\tau)(-dt^2 + \cosh^2t\;
d\Omega^2_{(3)})\, ,\ee
%%%%
which describes a multidimensional/multitemporal solution of the
Einstein equations. Relative to time $\tau$ we have the FRW-type
universe \rf{5.9} and with respect to time $t$ we obtain the de
Sitter universe with the Hubble constant $H=1$. The number of
spatial coordinates here is equal to usual three. In the present
paper we shall not investigate in detail solutions \rf{5.10} and
\rf{5.12} postponing their study for future work.

%%%%%%%%%%%%%%%%%%%%%%%%%%%%%%%%%%%%%%%%%%%%%%%%%%%%%%%%%%%%%%%%%%

\section{Other dust brane-world models\label{alternative}}
\setcounter{equation}{0}

\bigskip

%%%%%%%%%%%%%%%%%%%%
In previous sections we concentrated on the investigation of the
solution of Eqs.\rf{2.10} and \rf{2.12} in the case of dust
($\alpha=1$) when the bulk contains a negative cosmological
constant ($\Lambda_5 < 0$) and 4--D metric $g^{(4)}$ describes a
space with positive constant curvature  ($k=+1$). Let us
consider now the dust brane-world models for other combinations of
parameters $\Lambda_5$ and $k$.

\medskip

{\bf{ 1)$\;$ Negative bulk cosmological constant: $\Lambda_5 <0$}}

\medskip
{\it{ a)$\,$ Flat brane-world model: $k=0$}}

The solution for the Euclidean Eq.\rf{2.12} in this case is
%%%%%%%%
\be{6.1} a(r) = l (\Lambda \bar A^2)^{1/4} \left[\cosh
\frac{2r}{l}\right]^{1/2}\, ,\quad \qquad -\infty < r < +
\infty\ee
%%%%%
which describes a wormhole with metric \rf{2.11}. The constant of
integration is taken in such a way that $r=0$ corresponds to the
wormhole throat. To obtain a compact brane-world instanton we
suppose that $g^{(4)}$ is the compact metric.

The compactness of positive curvature spaces is evident. However,
Ricci-flat spaces and negative curvature spaces can be compact
too. This can be achieved by appropriate periodicity conditions
for the coordinates or, equivalently, through the action of
discrete groups $\Gamma $ of isometries related to face pairings
and manifold's topology \cite{Ellis}. The simplest example of Ricci-flat
compact spaces is given by $D$ - dimensional tori $T^D=\RR
^D/\Gamma $. In the case of negative curvature spaces,
d--dimensional spaces of constant negative curvature are isometric
to the open, simply connected, infinite hyperbolic space $H^d$.
But there exists also an infinite number of compact, multiply
connected, hyperbolic quotient manifolds $H^d/\Gamma $.

Thus, for simplicity we suppose that Ricci-flat space in our case
is 4--D torus. The brane world instantons can be obtained by the
same method as it was done in sections 3 with physically similar
properties described in section 4. Analytic continuation to the
Lorentzian region can be performed for any of coordinates of the
4--D torus. Then, we obtain a brane-world model with static flat
brane (3--D torus) or with a number of such parallel branes. A
multitemporal model can be also obtained here by analogy with the
construction of metric \rf{5.12}.

\medskip
{\it{ b)$\,$ Hyperbolic brane-world model: $k=-1$}}

In this case, solution of Eq.\rf{2.12} is
%%%%%%
\be{6.2} a(r) = \frac{l}{\sqrt{2}} \left( \sqrt{b}\, \cosh
\frac{2r}{l} +1 \right)^{1/2}\, ,\quad \qquad -\infty < r < +
\infty \, . \ee
%%%%
Metric \rf{2.11} with this solution again describes a wormhole.
The main difference between this model and the wormhole with the
scale factor \rf{2.14} consists in different 4--D metrics
$g^{(4)}$. In the case of solution \rf{6.2} $r$ constant slices
correspond to 4--D hyperbolic compact spaces $H^4/\Gamma$
%%%%
\be{6.3} g^{(4)}_{\mu \nu}dx^{\mu}dx^{\nu}= d\chi^2 +
\sinh^2\chi\, d\Omega^2_{(3)}\, .\ee
%%%%
The compact brane-world instantons can be constructed here by the
same procedure as that was used in section 3 and have the same
qualitatively physical properties as those described in section 4.
However, for metric \rf{6.3} branes are bent 4--D hyperbolic
(compact) spaces. It also follows from eq. \rf{6.3} that the
analytic continuation of the form \rf{5.2} to the Lorentzian
space-time is impossible here. However, we still can perform the
continuation $r \to i\tau$ to obtain a 5--D baby universe. Then,
brane-world models can be obtained by cutting and gluing the
compact hyperbolic space $H^4/\Gamma$ in the same way as it was
done for metric \rf{5.10} and scale factor \rf{5.11}.

For this case any multitemporal model is absent.
\medskip

{\bf{ 2)$\;$ Positive bulk cosmological constant: $\Lambda_5 >0$}}

\medskip
{\it{ a)$\,$ Positive curvature brane-world model: $k=+1$}}

It can be easily seen that for positive bulk cosmological
constant, Euclidean Eq.\rf{2.12} can have a solution only for
positive curvature of the 4--D space and this solution is a
periodic function
%%%%%%
\be{6.4} a(r) = \frac{l}{\sqrt{2}} \left( 1 - \sqrt{\bar b}\, \cos
\frac{2(r-r_{-})}{l}\right)^{1/2}\, , \qquad 2\pi n \le
\frac{2|r-r_{-}|}{l} \le (2n+1)\pi\, ,\quad n = 0,1,2,\ldots \,
,\ee
%%%%
where $r_{-}$ is an integration constant, $\bar b \equiv 1
-4\Lambda \bar A^2 > 0 \longrightarrow \bar A^2 < 1/(4\Lambda)$
and $\Lambda \equiv \Lambda_5/6$. Solution \rf{6.4} describes an
infinite string of wormholes with throats at $|r-r_{-}|/l = n\pi$
which can be smoothly glued at points of their maxima $|r-r_{-}|/l
= (2n+1)\pi /2$. Compact brane-world instantons can again be
constructed similarly to how we did in section 3. However, for
solution \rf{6.4} the sign of the brane tensions and positions of
the radial profile maxima depend on the choice of cutting. If we
cut wormholes before their maxima, we obtain the string of 4--D
spherical positive tension branes with the zero-mode localization
on them. But if we cut the wormholes after their maxima, then
branes will have negative tensions and zero-mode radial profile
will have local minima at these branes. So, we shall not consider
the latter type of branes . For the string of positive tension
branes, birth from "nothing" occurs exactly like it was described
in section 5.

However, in contrast to the negative cosmological bulk model, we
have now two types of 5--D baby universes. There are that branch
off from their throats where they get their maxima and reaches
singular minima at the bouncing points. Other baby universes
branch off from the maxima of the wormholes and have at that
moment non-singular minima. The latter type of baby universes
describes 5--D asymptotically de Sitter universes. For both types
we can construct also baby universe brane world models of the form
\rf{5.10}. There are two types of multitemporal models of the form
of Eq.\rf{5.12} depending on which of the baby universes
(described above) is taken for.

{\it{ b)$\,$ Flat 5--D Lorentzian model: $k=0$}}

In this case the Lorentzian solution of Eq. \rf{2.10} is
%%%%%
\be{6.5} a(\tau) = l (\Lambda \bar A^2)^{1/4} \left[\sinh
\frac{2|\tau -\tau_0|}{l}\right]^{1/2}\, ,\quad \qquad 0 \le |\tau
-\tau_0| < + \infty \, .\ee
%%%%%
It describes an asymptotically de Sitter 5--D universe, with
Ricci-flat 4--D $g^{(4)}$ metric. Obviously, it is impossible to
construct branes out from such a metric, but multitemporal models
of the type given by Eq.\rf{5.12} with flat Lorentzian 4--D metric
exist in this case.

{\it{ c)$\,$ Hyperbolic brane-world model: $k=-1$}}

Depending on the sign of $\tilde b \equiv -1 + 4\Lambda \bar A^2$,
there are 3 types of Lorentzian solutions of Eq.\rf{2.10},
%%%%%%
\be{6.6} a(\tau) =\left\{\begin{array}{rcl}
&\frac{l}{\sqrt{2}}\left\{ \sqrt{\tilde b}\, \sinh\, [\arcsinh
(1/\sqrt{\tilde b}) + 2|\tau -\tau_0|/l] - 1\right\}^{1/2}&\, ,
\quad \tilde b
> 0
\\ &\frac{l}{\sqrt{2}}\left[\exp (2|\tau - \tau_0|/l) -1\right]^{1/2}&\, ,
\quad \tilde b = 0
\\ &\frac{l}{\sqrt{2}}\left\{ \cosh (2|\tau -\tau_0|/l) + 2\sqrt{\Lambda
\bar A^2}\sinh (2|\tau - \tau_0|/l) -1 \right\}^{1/2}&\, , \quad
\tilde b < 0
\\
\end{array}\right. \ee
%%%%%
where $0\le |\tau - \tau_0| < +\infty$. All of these solutions
describe asymptotically de Sitter 5--D universes with 4--D
hyperbolic compact spaces $H^4/\Gamma$ \rf{6.3}. The brane-world
models for these solutions can be obtained by a similar procedure
to that was described for equations \rf{5.10} - \rf{5.11}. The
multitemporal models are absent in this case, c).

\medskip

{\bf{ 2)$\;$ Zero bulk cosmological constant: $\Lambda_5 = 0$}}

\medskip

In this case, the solutions of the Euclidean Eq.\rf{2.12} only
exist for positive curvature 4--D space $g^{(4)}$.

{\it{ a)$\,$ Positive curvature brane-world model: $k=+1$}}

The Euclidean solution is
%%%
\be{6.7} a(r) = \sqrt{\bar A^2 + r^2}\, ,\quad \qquad -\infty < r
< +\infty \, ,\ee
%%%%
which, describes a 5--D wormhole (the constant of integration
being again chosen in such a way that $r=0$ corresponds to the
wormhole throat). It is interesting to note that qualitatively the
same wormholes were obtained in papers of references
\cite{HL,Zhuk} for 4--D models with conformally coupled scalar
field (which is equivalent to radiation) and was used also in
paper \cite{Hawking} for investigations of wormholes.
Qualitatively the same instanton for 4--D models with axionic
field (which is equivalent to ultra-stiff matter) was found in
reference \cite{GS}.

Expression \rf{6.7} is qualitatively similar to solution \rf{2.14}. We can
construct the compact brane-world instantons with spherical branes
by the same procedure as it was used in section 3, with the same
physically properties as those described in sections 4 and 5.

{\it{ b)$\,$ Flat 5--D Lorentzian model: $k=0$}}

The Lorentzian solution of Eq.\rf{2.10} for this case reads:
%%%%
\be{6.8} a(\tau ) = \sqrt{2 \bar A |\tau - \tau_0|} \, , \quad
\qquad 0 \le |\tau - \tau_0| <+\infty\, .\ee
%%%%
It describe 5--D universes with FRW-like initial singularity. The
time derivative of the scale factor $da/d\tau \to 0$ when $|\tau|
\to \infty$. For this particular combination of parameters
$\Lambda_5$ and $k$ there is no brane world, though multitemporal
model analogous to that of solution \rf{5.12} can still be
obtained.

{\it{ c)$\,$ Hyperbolic brane-world model: $k=-1$}}

The Lorentzian solution of the Eq.\rf{2.10},
%%%%%
\be{6.9} a(\tau ) = \sqrt{-\bar A^2 + [|\bar A| + (\tau -\tau_0)]^2}\,
, \quad \qquad 0 \le |\tau -\tau_0| <+\infty\, ,\ee
%%%%%
describes asymptotically flat (Milne-type) 5--D hyperbolic
space-time. For small times this solution coincides with solution
\rf{6.8}. Brane-world models for solution \rf{6.9} can be
obtained by similar procedures to that used for models \rf{5.10}
and \rf{5.11}. Finally, multitemporal models of the type \rf{5.12}
are absent.

%%%%%%%%%%%%%%%%%%%%%%%%%%%%%%%%%%%%%%%%%%%%%%%%%%%%%%%%%%%%%%%%
\section{Conclusions}

\bigskip

In this paper we have considered in some detail 5--dimensional
cosmological models corresponding to the case of a bulk filled
with perfect fluid and a cosmological constant, particularizing to
the case of dust. Special attention has been given to the case of
a negative cosmological constant in a bulk with the geometry of a
four-sphere where we have found an asymptotically AdS wormhole
instantonic solution. Brane-world instantons with a single 4--D
spherical brane as well as with a string of such concentric branes
can then be built up by using a cutting and gluing procedure. We
have been able to obtain regular solutions which are free from any
singularities at the origin of extra coordinates, and can be
compactified so that the asymptotic divergences of the scale
factor are prevented. Zero-mode massless gravitons are shown to be
localized on these 4--D branes, so allowing such branes to nest
Newtonian gravity. Analytical continuation from the brane
instantonic metric to the Lorentzian regime leads to de Sitter
3--dimensional inflating branes. After birth of the inflating brane
world from "nothing", the perfect fluid (dust) remains frozen: it
is contained in the bulk but not on the brane. Here, inflation has
pure geometrical origin: the Hubble constant of the inflating 3--D
brane world is defined by the radius of the Euclidean 4--D
spherical brane.

Other kinds of brane world models were also obtained and
discussed, both for four-spheres and other spatial topologies.
Thus, we have found a class of brane world models characterized by
a common dynamical behaviour for extra-dimension and branes.

Of course, some important aspects of this research need further
investigations including transition from inflationary branes into
a matter-dominated brane universe, the physical consequences from the above
mentioned brane-world model with dynamical equivalence between extra
dimension and
branes, as well as the meaning of the 5-dimensional spacetimes
described by a metric with two timelike dimensions.

%%%%%%%%%%%%%%%%%%%%%%%%%%%%%%%%%%%%%%%%%%%%%%%%%%%%%%%%%%%%%%%%

\bigskip
{\bf Acknowledgments}

A.Z. thanks Instituto de Matem\'{a}ticas y F\'{\i}sica Fundamental, CSIC,
for kind hospitality during preparation of this paper. A.Z.
acknowledges support by Spanish Ministry of Education, Culture and
Sport (the programme for Sabbatical Stay in Spain) and the
programme SCOPES (Scientific co-operation between Eastern Europe
and Switzerland) of the Swiss National Science Foundation, project
No. 7SUPJ062239. M.B.L. is supported by a grant of the Spanish
Ministry of Science and Technology. This investigation was
supported by the GICYT under Research Project No. PB97-1218.

%%%%%%%%%%%%%%%%%%%%%%%%%%%%%%%%%%%%%%%%%%%%%%%%%%%%%%%%%%%%%%%%

%%%%%%%%%%%%%%%%%%%%%%%%%%%%%%%%%%%%%%%%%%%%%%%%%%%%%%%%%%%%%%%%

%%%%%%%%%%%%%%%%%%%%%%%%%%%%%%%%%%%%%%%%%%%%%%%%%%%%%%%%%%%%%%%%

\newpage

\noindent
\begin{picture}(10,8.4)
\epsfxsize=13cm
\epsfbox{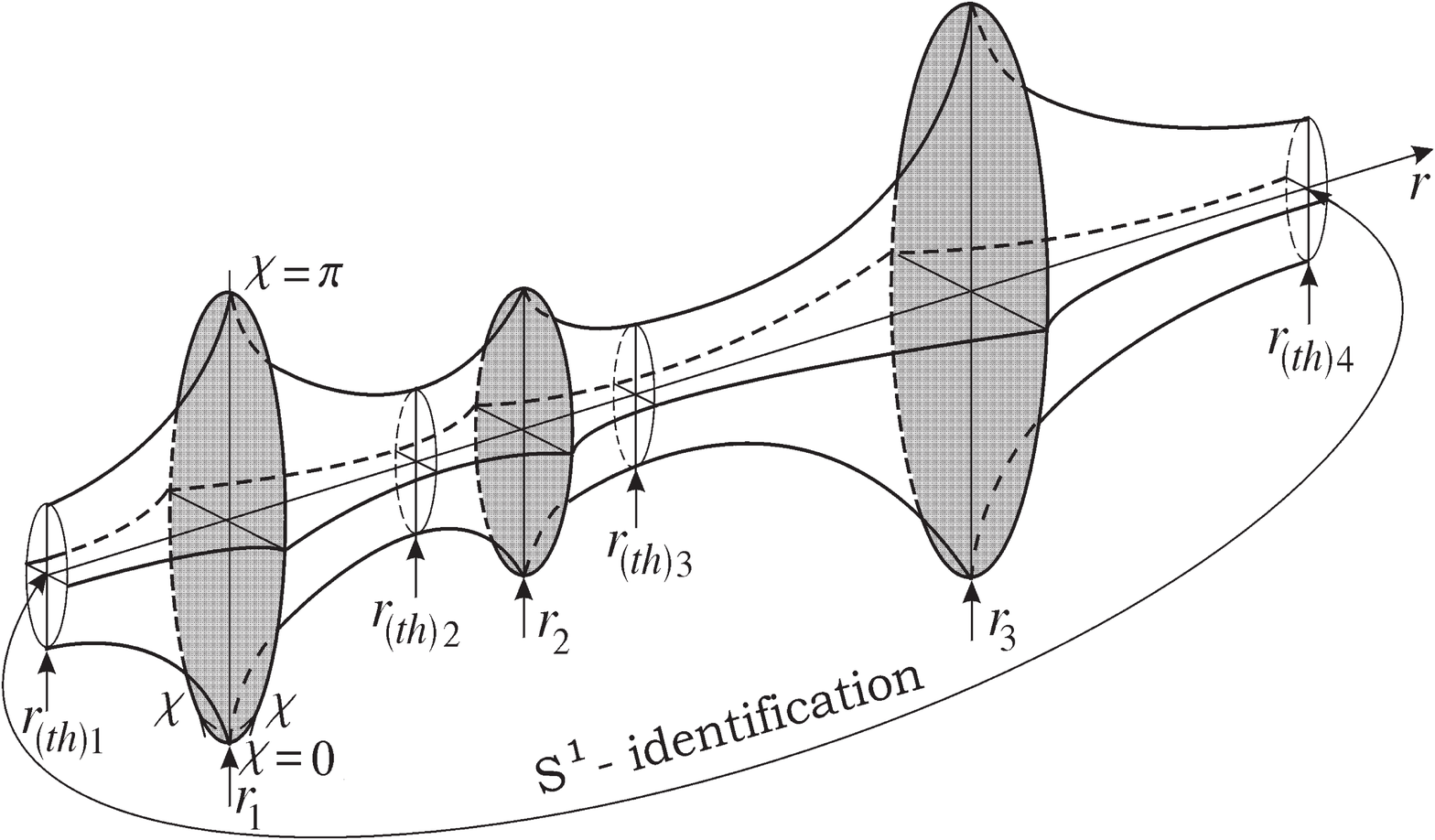}
%\put(0.5,8.9){\special{em:graph picture1.eps}}
\end{picture}

Figure 1: Compact n-brane-world instanton \rf{3.3} in the
case $n=3$. Each point in the figure represents a 3--D sphere. Lines with
fixed coordinates $r = \const \equiv r_0$ correspond to 4--D spheres with
radii $a(r_0)$. The branes are 4--D concentric spheres surrounding the 5--D
bulk.

\noindent
\begin{picture}(8,10.1)
\epsfxsize=13cm
\epsfbox{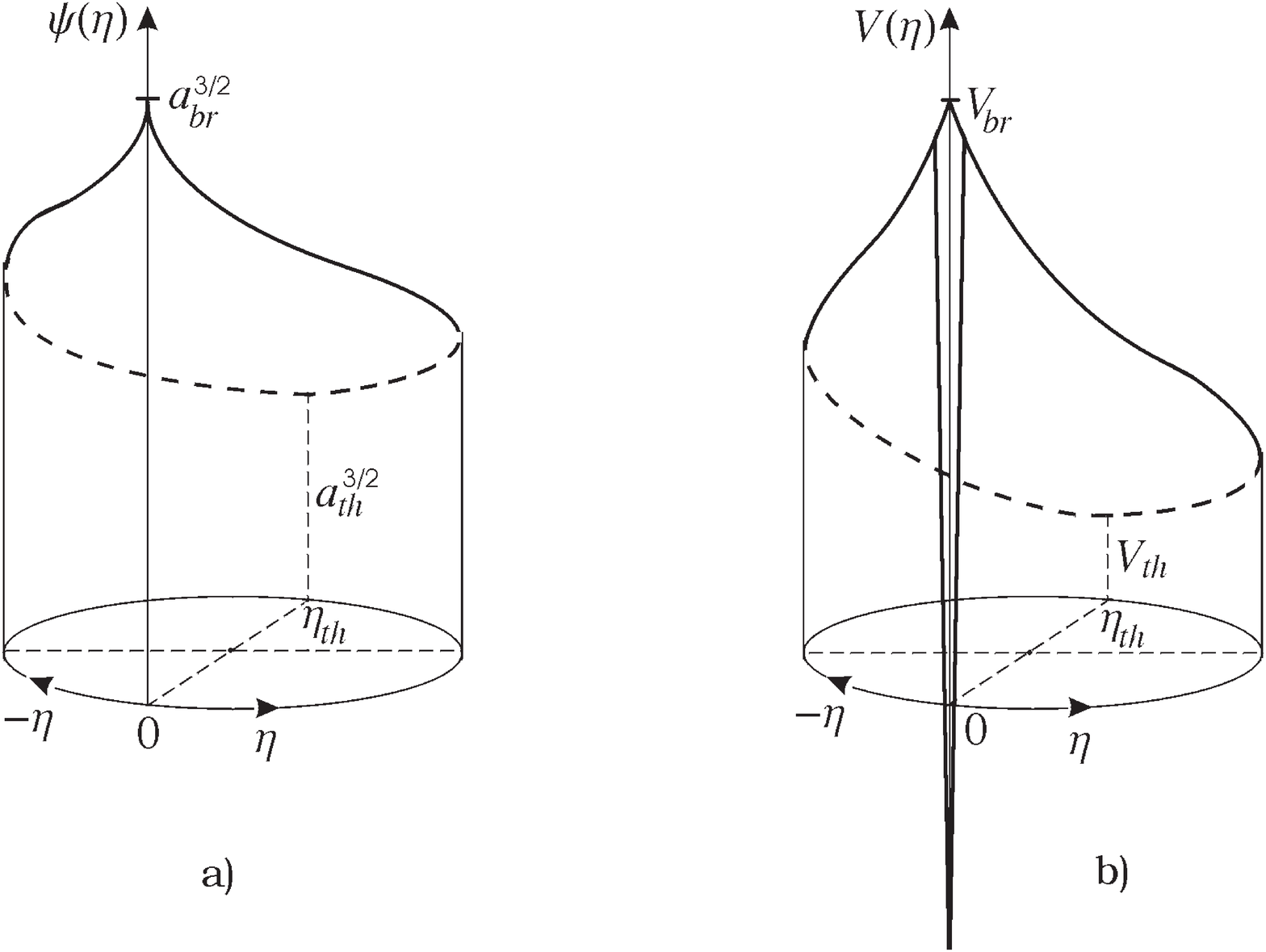}
%\put(1,9.4){\special{em:graph picture2.eps}}
%\put(7.5,10){\special{em:graph pic31.pcx}}
\end{picture}

Figure 2: (a) Zero-mode radial profile $\psi\sim a^{3/2}$
for the case of the compact brane--world instanton \rf{3.5}. $\psi$
has its maximum at the brane with coordinate $\eta =0$ and tends
to the minimum at the identified throats of wormholes at $\eta_{th}
=-b^{-1/4}K(m)+\eta_{0}\leftrightarrow b^{-1/4}K(m)-\eta_{0} $.

(b) Volcano potential \rf{4.4} for the compact
one-brane instanton \rf{3.5}. The wings of the potential evolves from its
maximum $V_{br}$ at the brane ($\eta =0$) to the minimum $V_{th}$ at the
identified throats ($\eta=\eta_{th}$). Mouth of the volcano (inverse
$\delta$--function) at $\eta =0$ provides the zero-mode localization on
the brane.

\newpage

\noindent
\begin{picture}(10,9.5)
\epsfxsize=11cm
\epsfbox{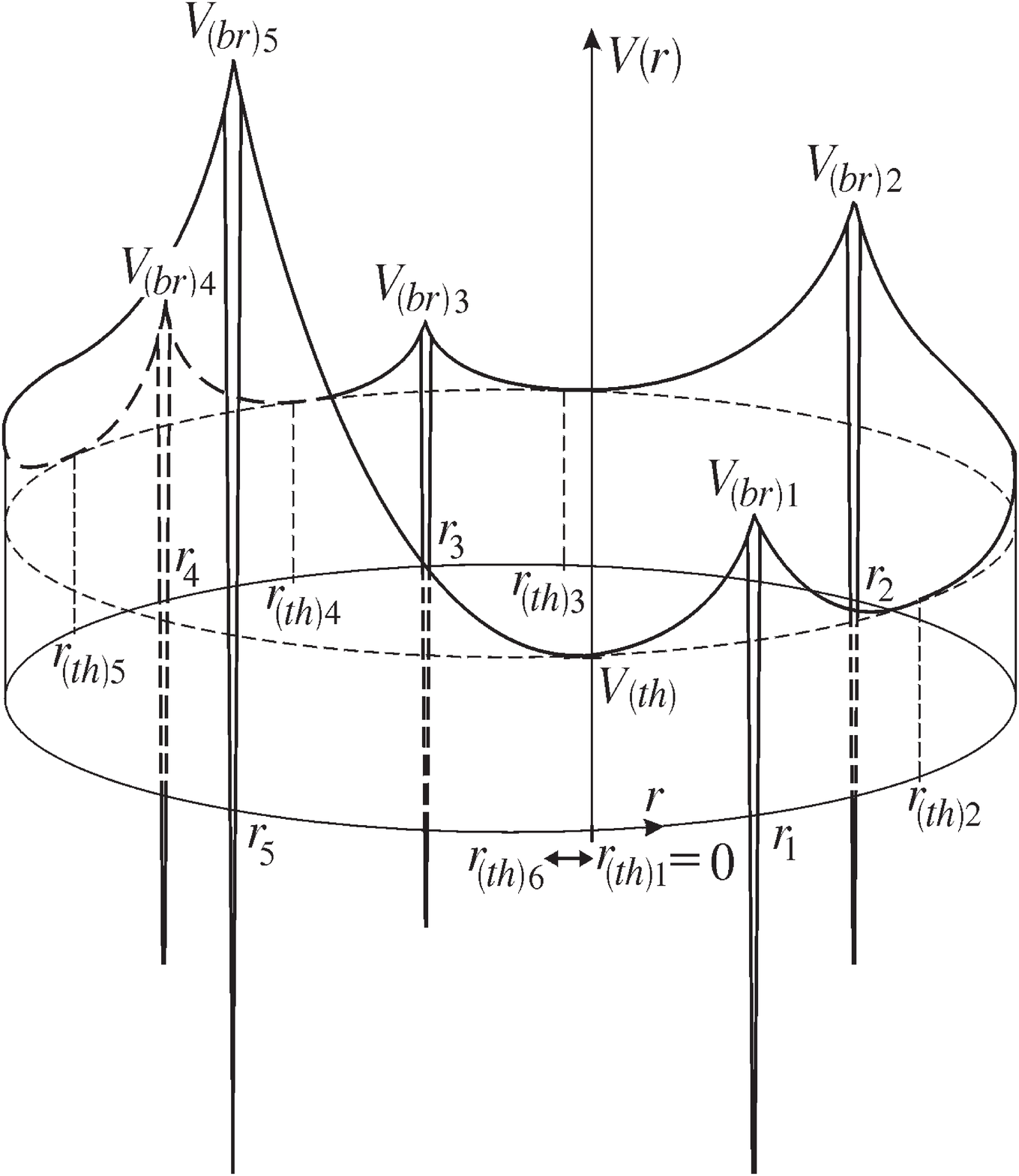}
%\put(2.5,10.5){\special{em:graph picture3.eps}}
\end{picture}

Figure 3: Kurile--ridge potential \rf{4.8} for the compact
5--brane-world instanton when $n=5$. Each of the "volcanoes" has a local
maximum $V_{(br)i}$ localized on the branes. Identical local minima
$V_{th}$ correspond to the wormhole throat positions $r_{(th)i}$.

\noindent
\begin{picture}(10,9.5)
\epsfxsize=13cm
\epsfbox{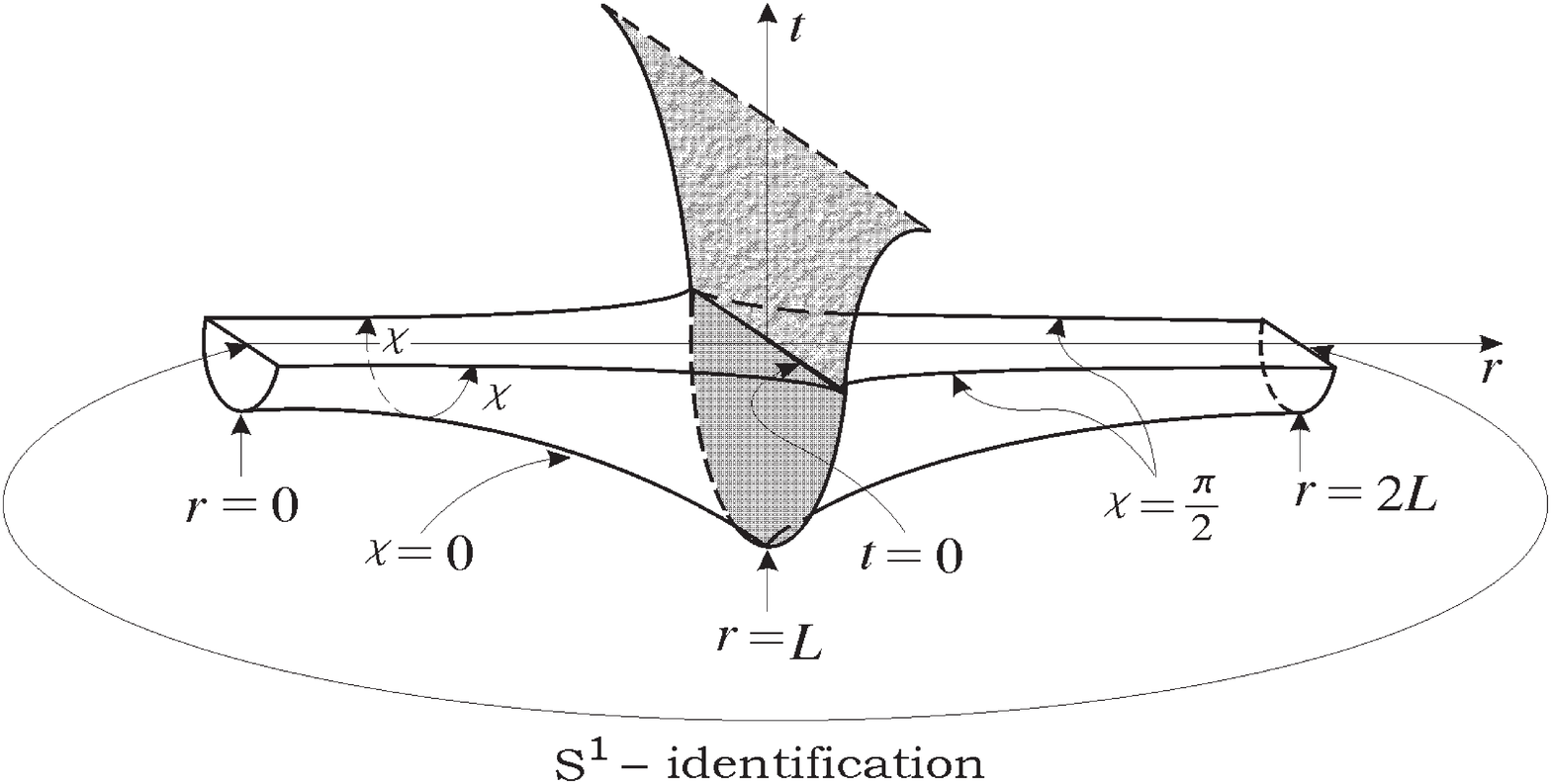}
%\put(1,8){\special{em:graph picture4.eps}}
\end{picture}

Figure 4: Birth of the one-brane-world $(n=1)$ from ``nothing''.
The creation takes place at the time $t=0$ and the 5--D Lorentzian metric is
described by Eq.\rf{5.3}. The brane after birth represents an inflating
3--D sphere with initial radius $H^{-1}$ that is de Sitter space-time
with Hubble constant $H^{-1}$.

\end{document}